\documentclass[twocolumn,aps,prl,draft]{revtex4}
\usepackage{graphicx}
\usepackage{bm}
\def\unitr{{\mbox{\boldmath$\hat r$}}}

\def\unitv{{\mbox{\boldmath$\hat v$}}} 
\def\f{{\mbox{\boldmath$f$}}}
\def\q{{\mbox{\boldmath$q$}}} 
\def\p{{\mbox{\boldmath$p$}}}
\def\k{{\mbox{\boldmath$k$}}}

\def\v{{\mbox{\boldmath$v$}}}

\def\sra {\rightarrow}
\newcommand{\x}{{\mbox{\boldmath$x$}}}

\newcommand{\be}{\begin{equation}} 
\newcommand{\ee}{\end{equation}}
\newcommand{\bea}{\begin{eqnarray}} 
\newcommand{\eea}{\end{eqnarray}}
\newcommand{\rv}{{\bf r}}

\newcommand{\la}{\left\langle}
\newcommand{\ra}{\right\rangle}
\begin{document} 
\title{Effects of forcing in three dimensional turbulent flows}
\author{Luca Biferale$^{1,4}$, Alessandra S.~Lanotte$^{2,4}$, and
Federico Toschi$^{3,4}$} \affiliation{$^1$ Dip. di Fisica,
Universit\`a "Tor Vergata", Via della Ricerca Scientifica 1, I-00133
Roma, Italy \\$^2$ Istituto per le Scienze dell'Atmosfera e del Clima,
CNR, Str.~Prov.~Lecce-Monteroni km.~1200, I-73100 Lecce, Italy \\$^3$
Istituto per le Applicazioni del Calcolo, CNR, Viale del Policlinico
137, I-00161 Roma, Italy \\$^4$ INFM, Unit\'a di Tor Vergata, I-00133
Roma, Italy}
\begin{abstract}
We present the results of a numerical investigation of
three-dimensional homogeneous and isotropic turbulence, stirred by a
random forcing with a power law spectrum, $E_f(k)\sim
k^{3-y}$. Numerical simulations are performed at different resolutions
up to $512^3$. We show that at varying the spectrum slope $y$,
small-scale turbulent fluctuations change from a {\it forcing
independent} to a {\it forcing dominated} statistics. We argue that
the critical value separating the two behaviours, in three dimensions,
is $y_c=4$. When the statistics is forcing dominated, for $y<y_c$, we
find dimensional scaling, i.e. intermittency is vanishingly small. On
the other hand, for $y>y_c$, we find the same anomalous scaling
measured in flows forced only at large scales. We connect these
results with the issue of {\it universality} in turbulent flows.
\end{abstract}
\maketitle
%%%%%%%%%%%%%%%%%%%%%%% 
The effects of both external forcing mechanisms and boundary
conditions on small-scale turbulent fluctuations have been the subject
of many theoretical, numerical and experimental studies
\cite{frisch,SA97}. The 1941 theory of Kolmogorov \cite{frisch} is
based on the assumption of {\it local} isotropy and homogeneity, that
is any turbulent flow, independently on the injection mechanism,
recovers universal statistical properties, for scales small enough
(and far from the boundaries). Indeed, experiments and numerical
simulations give strong indications that Eulerian and Lagrangian
isotropic/anisotropic small-scales velocity statistics are pretty
independent of the {\it large-scale} forcing mechanisms
\cite{SW02,GFN02,boden,BCTT03,BBCLTV03}. Still, we lack a firm
understanding for these evidences. From the theoretical point of view,
precious hints arise from {\it linear} problems, like passive scalar
or passively advected magnetic fields. For the class of Kraichnan
models \cite{kr94}, anomalous scaling has been shown to be associated
to statistically stationary solutions of the unforced equations for
correlation functions \cite{FGV01}. Scaling exponents are consequently
universal with respect of the injection mechanisms. Concerning
non-linear problems, as the Navier-Stokes case, analytical results
have been often pursued by means of the Renormalization Group (RG)
\cite{FNS77,FF78}. In the RG framework, turbulence is stirred at all scales
by a self-similar Gaussian field, with zero mean and white-noise in
time. The two-point correlation function in Fourier space is given by
\bea 
\label{def:force}
&\la f_i(\k,t) f_j(\k',t') \ra \propto& \\ & D_0 k^{4-d}(k_0^2 +
k^2)^{-y/2}P_{ij}(\k) \delta(\k+\k')\, \delta(t-t').& \nonumber
\eea 

Here $1/k_0 \sim L$ is the largest length in the system (infrared
cut-off), $D_0$ is the forcing intensity, $P_{ij}(\k)$ is the
projector assuring incompressibility and $d$ is the spatial dimension
(always assumed to be $d=3$ hereafter). The influence of the stirring
mechanism at small scales is governed by the value of the slope
$y$. We go from a situation when the forcing has a strong input at all
scales, $y \sim 0$\, originally investigated in \cite{FNS77}, to a
quasi {\it large-scale} forcing when $y \rightarrow
\infty$. Renormalization Group calculations, based on a $y$-expansion,
predict a power-law energy spectrum $E(k)\sim k^{1-2y/3}$, in the
domain $\eta \ll k^{-1} \ll L$, where $\eta$ is the viscous scale of
the system, and for $y \ll 1$. Notice that the Kolmogorov value,
$E(k)\sim k^{-5/3}$, describing experimental turbulent flows stirred
by a large-scale forcing, is obtained for $y=4$, i.e. quite far from
the perturbative region where the RG calculations are under
control. The Kolmogorov spectrum can be obtained, however, by means of
a simple dimensional analysis, still within the same framework
\cite{DDM79}. Extension of the RG formalism to finite $y$ values, up
to $y=4$, have been attempted with different kind of approximation
\cite{YO86,AAKV03} altough in a range where convergence of the RG
expansion is not granted anymore \cite{EY94}. As for the numerical
simulations, in \cite{SMP98} the problem has been investigated for
various $y$ values, and it has been shown that, for $y=4$, results are
in good agreement with the picture of large-scale forced turbulence,
while for $y<4$ the situation becomes less clear. However, the low
numerical resolution used in \cite{SMP98} makes these results far from
being conclusive.

Beside the issue connected to the RG approach, there exists a whole
set of interesting questions concerning turbulent flows with a
power-law forcing. To which extent small-scale fluctuations are
sensitive to the injection mechanism?  Does it exist a critical
value $y_c$ separating different regimes?  Can we observe anomalous
scaling in the forcing dominated case $y \sim 0$?  Let us suppose, for
example, that it exists a finite $y_c$, beyond which small-scale
statistics is forcing independent: this would rule out any attempt to
control intermittency analytically, by means of a perturbative
approach which starts from forcing dominated turbulent solutions at
$y\ll y_c$.

Hints on the problem can also come from the study of the one-dimensional
Burgers equation in presence of a power law forcing. In \cite{HJ97}, a
numerical study was presented showing that there is a critical value
of the forcing slope, such that the velocity field passes from the
usual bifractal statistics (observed in large-scale forced Burgers
flows), to a statistics influenced from the forcing. Surprisingly, also
in the forcing dominated regime, a non-trivial (multifractal?) scaling
was observed in \cite{HJ97}. A rigorous understanding of the mechanism
leading to this result is, however, still missing: we will come back to this
point later on, after having presented our results.

In the sequel, we address the problem of the small-scale statistical
properties of three dimensional turbulent flows at varying the
parameter $y$. We will show that at crossing $y_c=4$ there exist two
regimes for the velocity field statistics with well defined and
different scaling properties.

We solved the Navier-Stokes equations with a second-order hyperviscous
dissipative term $\propto \nu \Delta^2$. Temporal integration has been
carried over for about $20\div 30$ large-eddy turnover times. We
performed variuos experiments, at resolutions $128^3$, $256^3$ and
$512^3$, corresponding to a maximum Taylor's Reynolds number equal to
$Re_{\lambda}= 220$ for the $512^3$ run.  As for the stirring force,
we specialized in the two following cases, one for each regime: the
first with $y=3.5<y_c$, the second with $y=6>y_c$. We also show some
results obtained with an analityical large-scale forcing, i.e. the
equivalent of $y \sra \infty$. The range of the forcing, in Fourier
space, extends down to the maximum resolved wavenumber. As we are
always confined in a finite box, we neglect here possible subtle
behaviours due to infrared divergences in the injection mechanism.

We start considering what happens to the system when the slope of the
forcing is changed. It is instructive to consider the equation for
the energy flux through the wave\-num\-ber ${\it k}$: $\Pi({\it
k})\equiv - \pi {\it k}^2 \int \Im [ \la \,(\k \cdot
\unitv(\q))(\unitv(\k)\cdot \unitv(\p))\,\ra + \la\, (\k \cdot
\unitv(\p))(\unitv(\k)\cdot \unitv(\q))\,\ra ] {\mbox d}\p {\mbox
d}\q$, where the three wavevectors satisfy $\k +\p +\q =0$, and the
symbol $\Im$ stands for the imaginary part. Such equation is
equivalent to the K\'arm\'an-Howarth equation in physical space; it
states that in a stationary, isotropic and homogeneous flow, the
contribution to the energy flux $\Pi({\it k})$ due to the non-linear
terms balances the total energy input from the injection mechanism
(see \cite{my,frisch}):
\begin{equation}
\Pi({\it k}) \sim  \int_{k_0 < |{\k}| < \it k}
\Re(\la\f(\k)\v(-\k)\ra) {\mbox d}{\k},
\label{flusso}
\end{equation}
where $k_0 \sim 1/L$, the symbol $\Re$ stands for the real part and
where we have neglected dissipative effects. For the special class of
forcings (\ref{def:force}), the {\it rhs} of (\ref{flusso}) can be
further simplified to: $\Pi({\it k}) \sim \int_{k_0 < |\k| < {\it k}}
\la |\f(\k)|^2\ra {\mbox d}\k$. From (\ref{def:force}) the forcing
spectrum is $E_f(k)=\la |\f(\k,t)|^2 \ra \sim k^{3-y}$. It follows
that for $y \ge 4$, the energy flux is constant in Fourier space for
$kL \gg 1 $ (up to logarithmic corrections for $y=4$). In other words,
the energy injection is dominated by the small wave-number region in
the integral (\ref{flusso}). In this case we expect to be very close
to the typical experimental situation of turbulence with a
large-scale, analytical forcing: energy is transferred down-scale {\it
via } an intermittent cascade. Coherently, the third order
longitudinal structure function, $S^{(3)}(r) \equiv \la \left[(\v(\x +
\rv) - \v(\x))\cdot \unitr\right]^3 \ra$, follows a linear behaviour
in $r$ as predicted by the $4/5$ law \cite{frisch}. For $y<4$, the
energy flux no longer saturates to a constant value as a function of
${\it k}$. The integral in (\ref{flusso}) now becomes ultraviolet
dominated. The direct input of energy from the forcing mechanism
affects inertial range statistics in a self-similar way, down to the
smallest scales where dissipative terms start to be important. In this
situation, we get for the energy flux $\Pi({\it k}) \sim {\it
k}^{4-y}$, with a constant prefactor which depends on the ultraviolet
cut-off. The corresponding scaling behaviour for the third order
structure function is now given by $S^{(3)}(r)\sim r ^{y-3}$.

What about higher order statistics? One is tempted to guess that for
$y\ge 4$ the fluctuations induced by the injection mechanism are
always sub-leading, anomalous scaling being the result of the cascade
mechanism driven by the non-linear terms of the equations of
motion. If so, for $y > 4$ we should fall in the same class of
``universality'' of turbulence generated with large-scale forcing,
i.e. small-scale velocity fluctuations should be
universal. Therefore, as far as $y>4$, longitudinal structure
functions should scale as:
\begin{equation}
S^n(r) \equiv \la \left[(\v(\x + \rv) - \v(\x))\cdot \unitr\right]^n
\ra \sim r^{\zeta^{(n)}_{\infty}}.
\label{largescale}
\end{equation}
In (\ref{largescale}), we have denoted with $\zeta^n_{\infty}$ the
scaling exponents measured with a smooth large-scale forcing.

\begin{figure}[ht]
\includegraphics[draft=false, width=\hsize,keepaspectratio,
clip=true]{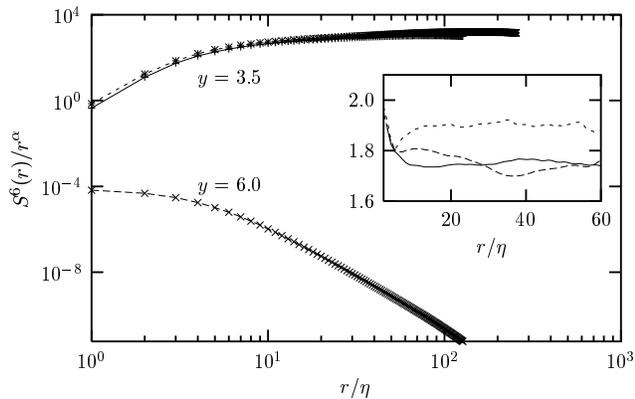}
\caption{Log-log plot of the compensated sixth-order structure
function $S^{(6)}(r)/r^{\alpha}$. The two top curves are for $y=3.5$
at the two resolutions $256^3$ and $512^3$: they are compensated with
the dimensional scaling \ref{smallscale}, i.e.  with an exponent
$\alpha=\zeta^{(6)}_{y=3.5}=0.5$. The bottom curve refers to the case
$y=6$, at the resolutions $256^3$, and is also compensated with the
exponent for the scaling (\ref{smallscale}),
$\alpha=\zeta^{(6)}_{y=6}=6$. Clearly the matching with the
dimensional exponent is not the correct one in the case $y=6$. Inset:
local slopes of the ESS curve for $S^{(6)}(r)$ {\it vs.} $S^{(3)}(r)$
at varying $r$. The top curve refers to the case $y=3.5$, and the two
bottom curves refer to the cases $y=6$ and $y \sra \infty$. The
dimensional scaling would correspond to the value $2$.}
\label{fig:1} 
\end{figure}

On the other hand, for $y < 4$, energy is directly injected in the
inertial range. Here, a dimensional matching with the forcing gives a
scaling behaviour which is always leading with respect to what
predicted in the $y>4$ range (\ref{largescale}).  We expect now that
anomalous scaling disappears, everything being dominated by the
Gaussian energy input at all scales. By the simple dimensional
argument connecting the scaling of structure functions to that of the
external forcing, for the range $y<4$ we have: \begin{equation}
S^{(n)}(r) \sim r^{\zeta^{(n)}_{y}}\,\,\, \mbox{with}\,\,
\zeta^{(n)}_{y}=\frac{n}{3}(y-3).
\label{smallscale}
\end{equation}
In Fig.~\ref{fig:1} we show the sixth order structure function,
$S^{(6)}(r)$ for the two cases $y=3.5$ and $y=6$, compensated with the
dimensional prediction given by the matching with the forcing
(\ref{smallscale}).
%----------------------------------------------------------------
%------------------------------------------------------------------ 
As it is clear, only for $y=3.5$ the statistics follows the forcing
injection obtaining a nice compensation. On the other hand, for $y=6$
the statistics is much closer to what usually measured with an
analytical large-scale forcing. This is quantitatively confirmed by
the inset picture, where we plot the logarithmic derivatives of
$S^{(6)}(r)$ {\it vs.} $S^{(3)}(r)$, for the two cases $y=3.5$ and
$y=6$, together with the results of the simulation with a large-scale,
analytical forcing corresponding to $y\sra \infty$. Here the local
slopes for $y=6$ and $y\sra \infty$ fluctuate around the same value,
compatible with those reported in literature \cite{SA97,GFN02}, while
the local slope for $y=3.5$ is different and tends to the expected
dimensional value. Values of all scaling exponents obtained in the
simulations are summarized in Table \ref{tabella}. Let us notice that
the measured exponents for the $y=3.5$ case are very close to the
non-intermittent, dimensional prediction. Only for high order moments
(i.e. $n=6$) there is a {\it small} deviation from the expected value
$\zeta^{(6)}/\zeta^{(3)}=2$.
%-----------------------------------------------------------------
\begin{table}[!b]
\begin{ruledtabular}
\begin{tabular}{|c|c|c|c|c|c|}
$n$ & 1 & 2 & 4 & 5 & 6 \\
\hline
$y_d$ & $0.333 $ & $0.666 $ & $1.33 $  & $1.66 $ & $2.00 $ \\ 
$y=3.5$        & $0.34 (1)$ & $0.67 (1)$ & $1.31 (2)$ & $1.62 (2)$ & $1.93 (3)$\\ 
$y=6$          & $0.36 (1)$ & $0.69 (1)$ & $1.28 (2)$ & $1.53 (3)$ & $1.75 (4)$ \\ 
$y\sra \infty$ & $0.36 (1)$ & $0.69 (1)$ & $1.27 (2)$ & $1.52 (3)$ & $1.75 (4)$ \\ 
\end{tabular}
\end{ruledtabular}
\caption{Scaling exponents in ESS, of the curves $S^{(n)}(r)$ {\it vs.}
$S^{(3)}(r)$, extracted from the following numerical simulations:
$y=3.5$, at resolution $256^3$ and $512^3$; $y=6$, at resolution
$256^3$; $y \sra \infty$, at resolution $512^3$. The first raw describes 
the dimensional values: $\zeta^{(n)}/\zeta^{(3)}=n/3$}
\label{tabella} 
\end{table}
%-----------------------------------------------------------------

\begin{figure}[!t]
\includegraphics[draft=false, width=\hsize,keepaspectratio,
clip=true]{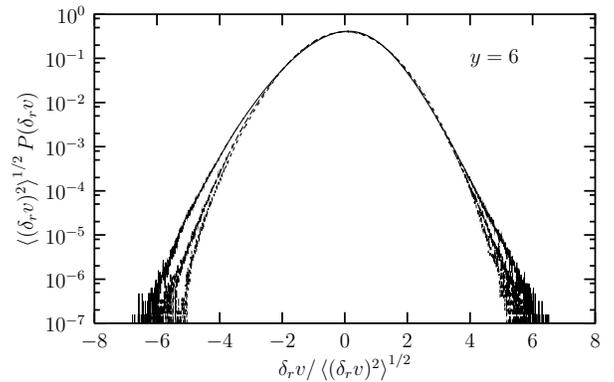}
\caption{PDF's of the velocity increments, for $y=6.0$, for three
separations $r_1=34\eta$ and $r_2=74\eta$ in
the inertial range, and $r_3=114\eta$ in the energy containing range.}
\label{fig:2} 
\end{figure}

\begin{figure}[!b] 
\includegraphics[draft=false, width=\hsize,keepaspectratio,
clip=true]{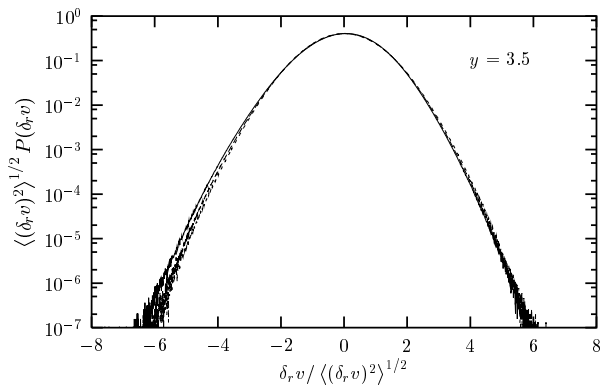}
\caption{PDF's of the velocity increments, for $y=3.5$. Scales
are the same of the previous picture: $r_1=34\eta$, $r_2=74\eta$, and
$r_3=114\eta$.}
\label{fig:3} 
\end{figure}

\begin{figure}[!t] 
\includegraphics[draft=false, width=\hsize,keepaspectratio,
clip=true]{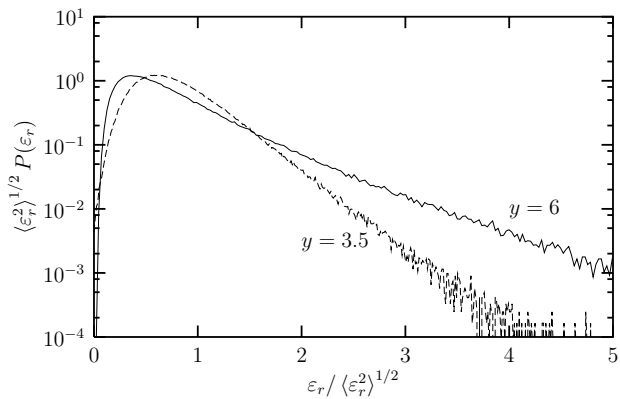}
\caption{PDF's of the coarse-grained energy dissipation ${\cal
P}(\varepsilon_r)$, for both $y=3.5$ and $y=6.0$, at the scale $r=8\eta$.}
\label{fig:4} 
\end{figure}

To quantify the level of intermittency at changing the scale, we also
plot the Probability Density Function (PDF) of velocity increments at
different scales, normalized to have unit variance. Figure \ref{fig:2}
shows the PDFs, in the case $y=6$, for three different separations
$r_1=34\eta$ and $r_2=74\eta$ in the inertial range, and $r_3=114\eta$
in the energy containing range. The three curves have larger tails
than a Gaussian distribution and have an intermittent behaviour, i.e.
they cannot be superposed. In Fig.~\ref{fig:3} we show the PDFs, for
the case $y=3.5$, at the same separations $(r_1,r_2,r_3)$. Now, the
three curves are almost indistinguishable, and show a very good
rescaling, a signature of the absence of intermittent effects. Only
for negative increments, a very tiny discrepancy is measured. It is
hard to say whether this is a robust effect or a spurious Reynolds
dependent phenomenon. We will come back to this issue later in the
conclusions.
%----------------------------------------------------------------
%------------------------------------------------------------------ 
%------------------------------------------------------------------ 
A dramatic difference at crossing the $y=y_c$ value is also observed
in the energy dissipation statistics. For both cases $y=3.5$ and
$y=6$, we measured the PDF's of the coarse-grained energy dissipation
$\varepsilon_r(\x)= V_r^{-1} \int_{V_r(\x)} \varepsilon(\x+\rv)
d{\rv}$, where $\varepsilon$ is the rate of dissipation for unit
volume, and the volume $V_r(\x)$ is centered at $\x$ and has
characteristic length scale $r \ll L$. In Fig.~\ref{fig:4} we compare
the PDF's ${\cal P}(\varepsilon_r)$ at the scale $r=8\eta$. Here the
results are even more impressive as the shape change is particularly
strong.

In this letter, we have presented clear evidences that turbulent
small-scale fluctuations, in presence of a direct injection of energy
at all scales, undergo a sharp transition for $y_c=4$. The first
regime, for $y >y_c$, is mainly forcing independent: small-scale
fluctuations develop anomalous scaling in agreement with what observed
in experiments and/or numerics obtained with a large-scale
forcing. This is a stringent test of turbulence universality: even if
directly affected by the injection of energy, small-scale fluctuations
show a robust behaviour. Things change abruptly at the critical value
of $y_c$, where the direct injection of energy becomes the dominant
effect in the inertial range. In this second regime, corresponding to
$y<y_c$, small-scale fluctuations get closer and closer to a Gaussian
statistics and intermittency disappear \cite{note}.
%----------------------------------------------------------------
%------------------------------------------------------------------ 
Before concluding, we discuss two possible mechanisms which could
partially disprove the last statement. First, even when $y<y_c$, we
may imagine that the intermittent energy cascade dominating the
statistics for $y>y_c$ might show up. For example, we may have that
for high order moments the forcing dominated solutions become
subleading with respect to those associated to the cascading
mechanism. In such case, the loss of rescaling in the PDF's tail of
Fig.~\ref{fig:2}, may be due to the survival of these rare anomalous
fluctuations. Second, even more complex is the scenario proposed in
\cite{HJ97}, where the possibility to have a forcing dependent
multiscaling statistics, when $y<y_c$, is conceived. This is not the
case for the linear Kraichnan models \cite{FGV01}, where forcing
dependent solutions are always dimensionally scaling. The main
difference is that, in the Navier-Stokes problem, the hierarchy of
equations for correlation function is unclosed: one cannot solve it
for a single order independently of all the others. In NS, being the
low order moments always forcing dominated for $y<y_c$, one may
observe some forcing dependency also on high order fluctuations, {\it
via} their coupling with low order correlation functions. This may be
a possible explanation of the multifractal behaviour observed in
Burgers equation \cite{HJ97}.\\

We are grateful to N.~V.~Antonov, J.~Bec, A.~Celani, U.~Frisch,
M.~Sbragaglia, and M.~Vergassola for useful discussions. This research
was supported in part by the EU under the Grants No. HPRN-CT
2000-00162 ``Non Ideal Turbulence'' and by the INFM (Iniziativa di
Calcolo Parallelo).
%%%%%%%%%%%%%%%%%%%%%%%%%%%%%%%%%% 
  

\begin{thebibliography}{99}
\bibitem{note} In the present work we refrained from investigating the case
with $y<3$. In this range velocity structure functions may become UV
divergent and important effects due to the matching with the
dissipative region may show up.
\end{thebibliography}
\end{document}